\documentclass[a4paper,10pt]{article}
\usepackage{amsmath,amsthm,amsfonts,graphicx}

\newcommand{\bra}[1]{\left\langle {#1} \right\vert}
\newcommand{\ket}[1]{\left\vert {#1} \right\rangle}
\newcommand{\R}{\mathbb{R}}
\newcommand{\C}{{\mathbb C}}
\newcommand{\ux}{\underline{x}}
\newcommand{\uy}{\underline{y}}

\newtheorem{theorem}{Theorem}[section]
\newtheorem{definition}{Definition}[section]
\newtheorem{problem}{Problem}[section]

\numberwithin{equation}{section}

\begin{document}

\title{\textbf{Towards a Theory of Conservative Computing}}
\author{{\sc G. Cattaneo$^\ast$, G. Della Vedova$^\dagger$, A. Leporati$^\ast$,
         R. Leporini\footnote{This work has been supported by MIUR$\backslash$%
         COFIN project ``Formal Languages and Automata: Theory and
         Applications''.}} \\[0.3cm]
    ${}^\ast$ Dipartimento di Informatica, Sistemistica e Comunicazione \\
    ${}^\dagger$ Dipartimento di Statistica \\
    Universit\`a degli Studi di Milano -- Bicocca \\
    Via Bicocca degli Arcimboldi 8, 20126 Milano, Italy \\[0.2cm]
    e-mail: \{ cattang, leporati, leporini \}@disco.unimib.it \\
    \hspace{-1.1cm}gianluca.dellavedova@unimib.it
 }
\date{}

\maketitle

\begin{abstract}
We extend the notion of conservativeness, given by Fredkin and Toffoli in 1982,
to generic gates whose input and output lines may assume a finite number $d$
of truth values.
A physical interpretation of conservativeness in terms of conservation of the
energy associated to the data used during the computation is given.
Moreover, we define \emph{conservative computations}, and we show that they
naturally induce a new NP--complete decision problem. Finally, we present a
framework that can be used to explicit the movement of energy occurring during
a computation, and we provide a quantum implementation of the primitives of
such framework using creation and annihilation operators on the Hilbert space
$\C^d$, where $d$ is the number of energy levels considered in the framework.
\end{abstract}

\section{Introduction}

Conservative logic has been introduced in \cite{fredkin-toffoli} as a
mathematical model that allows one to describe computations which reflect some
properties of microdynamical laws of Physics, such as reversibility and
conservation of the internal energy of the physical system used to perform the
computations.
The model is based upon the so called Fredkin gate, a
three--input/three--output Boolean gate originally introduced by Petri in
\cite{petri}, whose input/output map $\text{\sc FG} : {\{0,1\}^3 \rightarrow
\{0,1\}^3}$ associates any input triple $(x_1,x_2,x_3)$ with its corresponding
output triple $(y_1,y_2,y_3)$ as follows:
\[
   y_1 = x_1
   \hspace{0.7cm}
   y_2 = (\lnot x_1\land x_2) \lor (x_1\land x_3)
   \hspace{0.7cm}
   y_3 = (x_1 \land x_2) \lor (\lnot x_1 \land x_3)
\]

The Fredkin gate is \emph{functionally complete} for the Boolean logic: by
fixing $x_3 = 0$ we get $y_3 = x_1 \land x_2$, whereas by fixing $x_2 = 1$ and
$x_3 = 0$ we get $y_2 = \lnot x_1$.
A useful point of view is that the Fredkin gate behaves as a \emph{conditional
switch}: that is, $\text{\sc FG}(1, x_2, x_3) = (1, x_3, x_2)$ and
$\text{\sc FG}(0, x_2, x_3) = (0, x_2, x_3)$ for every $x_2, x_3 \in \{0,1\}$.
In other words, the first input line can be considered as a control line whose
value determines whether the input values $x_2$ and $x_3$ have to be exchanged
or not.

According to \cite{fredkin-toffoli}, \emph{conservativeness} is usually modeled
by the property that the output patterns of the involved gates are always a
permutation of the patterns given in input.
Let us stress that this does not mean that a fixed permutation is applied to
every possible input pattern; on the contrary, the applied permutation depends
on the input pattern.
Here we just mention the fact that every permutation can be written as a
composition of transpositions.
Hence not only the Fredkin gate can be used to build an appropriate circuit to
perform any given conservative computation (and thus it is universal also in
this sense with respect to conservative computations), but it is also the most
elementary conceivable operation that can be used to describe conservative
computations.
In this paper we will propose some analogous elementary operations with respect
to our notion of conservativeness.

The Fredkin gate is also \emph{reversible}, that is, {\sc FG} is a bijective
map on $\{0,1\}^3$.
Notice that conservativeness and reversibility are two independent notions: a
gate can satisfy both properties, only one of them, or none.
Since every reversible gate computes a bijective map between input and output
patterns, and every conservative gate produces permutations of its input
patterns, it follows that they must necessarily have the same number of input
and output lines.

In this paper we extend the notion of conservativeness to generic gates whose
input and output lines may assume a finite number $d$ of truth values, and we
derive some properties which are satisfied by conservative gates.
By associating equispaced energy levels to the truth values, we show that our
notion of conservativeness corresponds to the energy conservation principle
applied to the data which are manipulated during the computation.
Let us stress that we are \emph{not} saying that the entire energy used to
perform the computation is preserved, or that the computing device is a
conservative physical system.
In particular we do not consider the energy needed to transform the input
values into output values, that is, the energy needed to \emph{perform} the
computation.

Successively we introduce the notion of \emph{conservative computation}, based
upon gates which are able to store some finite amount of energy and to reuse
it during the computation.
We show that the decision problem to determine whether a given computation can
be performed in a conservative way through a gate which is able to store at
most $C$ units of energy is NP--complete.

Finally, we introduce a framework that allows one to visualize the movement of
energy occurring during a computation performed by a generic gate.
The framework is based upon some primitive operators that conditionally move
one unit of energy between any two given input/output lines of the gate.
Using creation and annihilation operators on the Hilbert space $\C^d$, we show
a quantum realization of these non--unitary conditional movement operators.

\section{Conservativeness}

Our notion of conservativeness, and the framework we will introduce, are based
upon many--valued logics.
These are extensions of the classical Boolean logic which are widely used to
manage incomplete and/or uncertain knowledge.
Different approaches to many--valued logics have been considered in literature:
for an overview, see \cite{rescher, rosser-turquette}.
However, here we are not interested into the study of syntactical or algebraic
aspects of many--valued logics; we just define some gates whose input and
output lines may assume ``intermediate'' truth values, such as the gates
defined in \cite{cattaneo-leporati-leporini-mv}.

For every integer $d \ge 2$, we consider the finite set $L_d = \{ 0,
\frac{1}{d-1}, \frac{2}{d-1}, \ldots$, $\frac{d-2}{d-1}, 1\}$ of truth values;
$0$ and $1$ denote falsity and truth, respectively, whereas the other values of
$L_d$ indicate different degrees of indefiniteness.
As usually found in literature, we will use $L_d$ both as a set of truth
values and as a numerical set equipped with the standard order relation on
rational numbers.

An $n$--input/$m$--output $d$--valued \emph{function} (also called an $(n, m,
d)$--function for short) is a map $f: L_d^n \to L_d^m$.
Analogously, an $(n, m, d)$--\emph{gate} and an $(n, m, d)$--\emph{circuit}
are devices that compute $(n, m, d)$--functions.
A gate is considered as a \emph{primitive operation}, that is, it is assumed
that a gate cannot be decomposed into simpler parts.
On the other hand, a circuit is composed by \emph{layers} of gates, where any
two gates $G_1$ and $G_2$ of the same layer satisfy the property that no
output line of $G_1$ is connected to any input line of $G_2$.

Let us consider the set ${\cal E}_d = \left\{ \varepsilon_0,
\varepsilon_{\frac{1}{d-1}}, \varepsilon_{\frac{2}{d-1}}, \ldots,
\varepsilon_{\frac{d-2}{d-1}}, \varepsilon_1 \right\} \subseteq \R$ of real
values; for exposition convenience, we can think to such quantities as energy
values.
To each truth value $v \in L_d$ we associate the energy level $\varepsilon_v$;
moreover, let us assume that the values of ${\cal E}_d$ are all positive,
equispaced, and ordered according to the corresponding truth values:
$0 < \varepsilon_0 < \varepsilon_{\frac{1}{d-1}} < \cdots <
\varepsilon_{\frac{d-2}{d-1}} < \varepsilon_1$.
If we denote by $\delta$ the gap between two adjacent energy levels then the
following holds:
\begin{equation}
   \varepsilon_v = \varepsilon_0 + \delta \, (d-1) \, v
   \hspace{1.5cm}
   \forall \, v \in L_d
   \label{eq:linear-relation}
\end{equation}
Notice that it is not required that $\varepsilon_0 = \delta$.

Now, let $\ux = (x_1, \ldots, x_n) \in L_d^n$ be an input \emph{pattern} for an
$(n, m, d)$--gate.
We define the \emph{amount of energy associated to $\ux$} as $E_n(\ux) =
\sum_{i=1}^n \varepsilon_{x_i}$, where $\varepsilon_{x_i} \in {\cal E}_d$ is
the amount of energy associated to the $i$--th element $x_i$ of the input
pattern.
Let us remark that the map $E_n : L_d^n \to \R^+$ is indeed a family of
mappings parameterized by $n$, the size of the input. Analogously, for an
output pattern $\uy \in L_d^m$ we define the associated amount of energy as
$E_m(\uy) = \sum_{i=1}^m \varepsilon_{y_i}$.
We can now define a conservative gate as follows.

\begin{definition}
   An $(n, m , d)$--gate, described by the function $G: L_d^n \to L_d^m$, is
   \emph{conservative} if the following condition holds:
   \begin{equation}
      \forall \, \ux \in L_d^n \qquad E_n(\ux) = E_m(G(\ux))
      \label{eq:conservativegate}
   \end{equation}
\end{definition}

Notice that it is not required that the gate has the same number of input and
output lines, as it happens with the reversible and conservative gates
considered in \cite{fredkin-toffoli, cattaneo-leporati-leporini-mv,
cattaneo-leporati-leporini-qm}.

Using relation \eqref{eq:linear-relation}, equation \eqref{eq:conservativegate}
can also be written as:
\[
   \frac{\varepsilon_0 n}{\delta (d-1)} + \sum_{i=1}^n x_i =
      \frac{\varepsilon_0 m}{\delta (d-1)} + \sum_{j=1}^m y_j
\]
Hence, when $n = m$ (as it happens, for example, with reversible gates)
conservativeness reduces to the conservation of the sum of truth values given
in input, as in weak conservativeness introduced in
\cite{cattaneo-leporati-leporini-mv}.
In the Boolean case this is equivalent to requiring that the number of $1$'s
given in input is preserved, as in the original notion of conservativeness
given in \cite{fredkin-toffoli}.

An interesting remark is that conservativeness entails an upper and a lower
bound to the ratio $\frac{m}{n}$ of the number of output lines versus the
number of input lines of a gate.
In fact, the maximum amount of energy that can be associated to an input
pattern is $\sum_{i=1}^n \varepsilon_1 = n \, \varepsilon_1$, whereas the
minimum amount of energy that can be associated to an output pattern is
$\sum_{i=1}^m \varepsilon_0 = m \, \varepsilon_0$.
Clearly, if it holds $n \, \varepsilon_1 < m \, \varepsilon_0$ then the gate
cannot produce any output pattern in a conservative way.
As a consequence, it must hold $\frac{m}{n} \le
\frac{\varepsilon_1}{\varepsilon_0}$.
Analogously, if we consider the minimum amount of energy $n \, \varepsilon_0$
that can be associated to an input pattern $\ux$ and the maximum amount of
energy $m \, \varepsilon_1$ that can be associated to an output pattern $\uy$,
it clearly must hold $n \, \varepsilon_0 \le m \, \varepsilon_1$, that is
$\frac{m}{n} \ge \frac{\varepsilon_0}{\varepsilon_1}$.
Summarizing, we have the bounds $\frac{\varepsilon_0}{\varepsilon_1} \le
\frac{m}{n} \le \frac{\varepsilon_1}{\varepsilon_0}$, that is, for a
conservative gate (or circuit) the number $m$ of output lines is constrained to
grow linearly with respect to the number $n$ of input lines.

A natural question is whether we can compute all functions in a conservative
way.
Let us consider the Boolean case.
Let $f: \{0,1\}^n \to \{0,1\}^m$ be a non necessarily conservative function,
and let us define the following quantities:
\begin{align*}
   &O_f = \max\left\{0, \max\limits_{\ux \in \{0,1\}^n} \left\{E_m(f(\ux)) -
           E_n(\ux)\right\}\right\} \\
   &Z_f = \max\left\{0, \max\limits_{\ux \in \{0,1\}^n} \left\{E_n(\ux) -
              E_m(f(\ux))\right\}\right\}
\end{align*}
Informally, $O_f$ (resp., $Z_f$) is the maximum number of $1$'s (resp., $0$'s)
in the output pattern that should be converted to $0$ (resp., $1$) in order to
make the computation conservative.
This means that if we use a gate $G_f$ with $n + O_f + Z_f$ input lines and
$m + O_f + Z_f$ output lines then we can compute $f$ in a conservative way as
follows:
\[
   G_f(\ux, \underline{1}_{O_f}, \underline{0}_{Z_f}) =
      (f(\ux), \underline{1}_{w(\ux)}, \underline{0}_{z(\ux)})
\]
where $\underline 1_k$ (resp., $\underline 0_k$) is the $k$--tuple consisting
of all $1$'s (resp., $0$'s), and the pair $(\underline{1}_{w(\ux)},
\underline{0}_{z(\ux)}) \in \{0,1\}^{O_f + Z_f}$ is such that $w(\ux) = {O_f +
E_n(\ux) - E_m(f(\ux))}$ and $z(\ux) = Z_f - E_n(\ux) + E_m(f(\ux))$.

As we can see, we use some additional input (resp., output) lines in order to
provide (resp., remove) the required (resp., exceeding) energy that allows
$G_f$ to compute $f$ in a conservative way.
It is easy to see that the same trick can be applied to generic $d$--valued
functions $f: L_d^n \to L_d^m$; instead of the number of missing or exceeding
$1$'s, we just compute the missing or exceeding number of energy units, and we
provide an appropriate number of additional input and output lines.

\section{Conservative computations}

Let us now introduce the notion of \emph{conservative computation}.
Let $G: L_d^n \to L_d^m$ be the function computed by an $(n, m, d)$--gate.
Moreover, let $S_{in} = \langle \ux_1, \ux_2, \ldots$, $\ux_k \rangle$ be a
sequence of elements from $L_d^n$ to be used as input patterns for the gate,
and let $S_{out} = \langle G(\ux_1), G(\ux_2), \ldots, G(\ux_k) \rangle$ be the
corresponding sequence of output patterns from $L_d^m$.
Let us consider the quantities $e_i = E_n(\ux_i) - E_m(G(\ux_i))$ for all
$i \in \{1, 2, \ldots, k\}$; note that, without loss of generality, by an
appropriate rescaling we may assume that all $e_i$'s are integer values.
We say that the \emph{computation} of $S_{out}$, obtained starting from
$S_{in}$, is \emph{conservative} if the following condition holds:
\[
   \sum_{i=1}^k e_i
       = \sum_{i=1}^k E_n(\ux_i) - \sum_{i=1}^k E_m(G(\ux_i)) = 0
\]
This condition formalizes the requirement that the total energy provided by
\emph{all} input patterns of $S_{in}$ is used to build all output patterns of
$S_{out}$.
Of course it may happen that $e_i > 0$ or $e_i < 0$ for some $i \in \{1, 2,
\ldots, k\}$.
In the former case the gate has an excess of energy that should be dissipated
into the environment after the production of the value $G(\ux_i)$, whereas in
the latter case the gate does not have enough energy to produce the desired
output pattern.
Since we want to avoid these situations, we assume to perform computations
through gates which are equipped with an internal \emph{accumulator} (also
\emph{storage unit}) which is able to store a maximum amount $C$ of energy
units.
We call $C$ the \emph{capacity} of the gate.
The amount of energy contained into the internal storage unit at a given time
can thus be used during the next computation step if the energy of the output
pattern that must be produced is greater than the energy of the corresponding
input pattern.

If the output patterns $G(\ux_1)$, $G(\ux_2), \ldots, G(\ux_k)$ are computed
exactly in this order then, assuming that the computation starts with no energy
stored into the gate, it is not difficult to see that  $\;st_1 := e_1, \;
st_2 := e_1 + e_2, \; \ldots, \; st_k := e_1 + e_2 + \ldots + e_k$ is the
\emph{sequence of the amounts of energy stored} into the gate during the
computation of $S_{out}$.
We say that a given conservative computation is \emph{$C$--feasible} if $0 \le
st_i \le C$ for all $i \in \{1, 2, \ldots, k\}$.
Notice that for conservative computations it always holds $st_k = 0$.

In some cases the order with which the output patterns of $S_{out}$ are
computed does not matter.
We can thus consider the following problem: Given an $(n, m, d)$--gate that
computes the map $G: L_d^n \to L_d^m$, an input sequence $\ux_1, \ldots, \ux_k$
and the corresponding output sequence $G(\ux_1), \ldots$, $G(\ux_k)$, is there
a permutation $\pi \in S_k$ (the symmetrical group of order $k$) such that the
computation of $G(\ux_{\pi(1)}), G(\ux_{\pi(2)}), \ldots, G(\ux_{\pi(k)})$ is
$C$--feasible?
This is a decision problem, whose relevant information is entirely provided by
the values $e_1, e_2, \ldots, e_k$, which can be formally stated as follows.

\begin{problem}
   {\sc Name}: {\sc ConsComp}.
   \begin{itemize}
      \item {\sc Instance}: a set ${\cal E} = \{e_1, e_2, \ldots, e_k\}$ of
            integer numbers such that $e_1 + e_2 + \ldots + e_k = 0$, and an
            integer number $C > 0$.
      \item {\sc Question}: is there a permutation $\pi \in S_k$ such that
            $\forall \, i \in \{1, 2, \ldots, k\}$
            \begin{equation}
               0 \le \sum_{j=1}^i e_{\pi(j)} \le C \quad ?
               \label{eq:constraints}
            \end{equation}
   \end{itemize}
\end{problem}

The {\sc ConsComp} problem can be obviously solved by trying every possible
permutation $\pi$ from $S_k$.
However, this procedure requires an exponential time with respect to $k$, the
length of the computation.
A natural question is whether it is possible to give the correct answer in
polynomial time.
With the following theorem we show that the {\sc ConsComp} problem is
NP--complete.
As it is well known \cite{garey-johnson}, this means that if there would exist
a polynomial time algorithm that solves the problem then we could immediately
conclude that the two complexity classes P and NP coincide, a very unlikely
situation.

\begin{theorem}
   {\sc ConsComp} is {\rm NP}--complete.
\label{teo:CONSCOMP}
\end{theorem}
\begin{proof}
{\sc ConsComp} is clearly in NP, since a permutation $\pi \in S_k$ has linear
length and verifying whether $\pi$ is a solution can be done in polynomial
time.
In order to conclude that {\sc ConsComp} is NP--complete, let us show a
polynomial reduction from {\sc Partition}, which is a well known NP--complete
problem \cite[page 47]{garey-johnson}.

Let $A = \{a_1, a_2, \ldots, a_k\}$ be a set of positive integer numbers, and
let $m = \sum_{i=1}^k a_i$.
The set $A$ is a positive instance of {\sc Partition} if and only if there
exists a set $A' \subseteq A$ such that $\sum_{a \in A'} a = \frac{m}{2}$.
If $m$ is odd then $A$ is certainly a negative instance, and we can associate
it to any negative instance of {\sc ConsComp}.
On the other hand, if $m$ is even we build the corresponding instance
$({\cal E}, C)$ of {\sc ConsComp} by putting $C = \frac{m}{2}$ and  ${\cal E}
= \{e_1, e_2, \ldots, e_k, e_{k+1}, e_{k+2}\}$, where $e_i = -a_i$ for all
$i \in \{1,2,\ldots,k\}$ and $e_{k+1} = e_{k+2} = \frac{m}{2}$.
It is immediately seen that this construction can be performed in polynomial
time.

We claim that $A$ is a positive instance of {\sc Partition} if and only if
$({\cal E}, C)$ is a positive instance of {\sc ConsComp}.
In fact, let us assume that $A$ is a positive instance of {\sc Partition}.
Then there exists a set $A' \subseteq A$ such that $\sum_{a \in A'} a =
\frac{m}{2}$, and the corresponding negative elements of ${\cal E}$ constitute
a subset ${\cal E}'$ such that $\sum_{e \in {\cal E}'} e = -\frac{m}{2}$.
We build a permutation $\pi \in S_k$ by selecting first the element $e_{k+1}$
followed by the elements of ${\cal E}'$ (chosen with any order), and then
$e_{k+2}$ followed by the remaining elements of ${\cal E}$.
It is immediately seen that $\pi$ satisfies the inequalities stated in
\eqref{eq:constraints}, and hence $({\cal E}, C)$ is a positive instance of
{\sc ConsComp}.
Conversely, let us assume that $({\cal E}, C)$ is a positive instance of
{\sc ConsComp}.
Then there exists a permutation $\pi \in S_k$ that satisfies the inequalities
stated in \eqref{eq:constraints}.
Since the first chosen element cannot be negative, it must necessarily be
$\frac{m}{2}$.
Moreover, since $C = \frac{m}{2}$, the second $\frac{m}{2}$ can be chosen if
and only if the energy stored into the gate is zero, that is, if and only if
there exists a set ${\cal E}' \subseteq {\cal E}$ of negative elements whose
sum is equal to $-\frac{m}{2}$.
The opposites of these elements constitute a set $A' \subseteq A$ such that
$\sum_{a \in A'} a = \frac{m}{2}$, and thus we can conclude that $A$ is a
positive instance of {\sc Partition}.
\end{proof}

\section{A framework for the study of energy--based properties of computations}

In this section we introduce a framework which can be used to define and study
energy--based properties of computations performed by $(n,m,d)$--gates.
The crucial idea of our framework is that we look at computations as a sequence
of \emph{conditional movements} of energy.
That is, the gate computes its output pattern as follows: for a given subset of
input lines, a condition on their values is checked; if this condition is
verified then a given action is performed, transforming such values, otherwise
no transformation is applied.
Successively, another condition is checked on another subset of lines
(comprising the output lines from the first step of computation), which
determines whether another action has to be performed, and so on until the
required values are obtained on the output lines.

To realize the gate according to the above procedure, we need a (Boolean)
\emph{control equipment}, and two \emph{primitives} to conditionally move
energy from a given line to another one.
We call these primitives \emph{conditional up} ({\sc CUp}) and
\emph{conditional down} ({\sc CDown}).
The realization of the gate can thus be viewed as a circuit composed by these
simpler elements.
Let us first describe {\sc CUp} and {\sc CDown} as $d$--valued gates.
In the following, we will provide a quantum realization as formulas composed of
creation and annihilation operators on $\C^d$, as we have done for the gates
presented in \cite{cattaneo-leporati-leporini-qm}.

\begin{figure}[t]
\begin{center}
   \includegraphics[width=12cm]{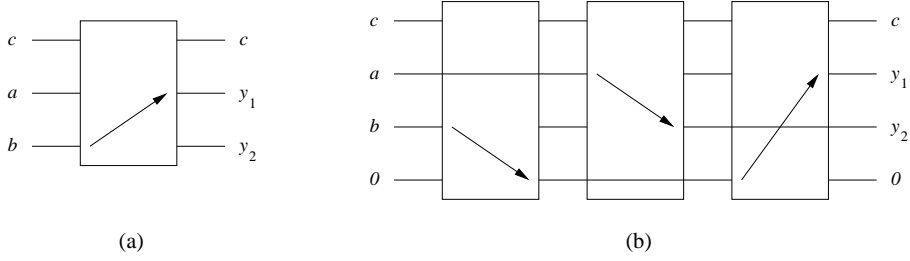}
\end{center}
\vspace{-0.4cm}
\caption{\small\it (a) The Conditional Up {\sc (CUp)} gate. (b) Realization of
         the Boolean Fredkin gate through two--valued {\sc CUp}'s and
         {\sc CDown}'s.}
\label{fig:figures}
\end{figure}

The {\sc CUp} gate is depicted in Figure \ref{fig:figures} (a).
It is a $(3,3,d)$--gate whose behavior is:

\medskip
\hspace{1cm}{\bf Input}: $(c, a, b) \in L_d^3$

\hspace{1cm}{\tt if} $c = 1$

\hspace{1.4cm} {\tt then} {\bf Output} $(c, \; a + \frac{1}{d-1}, \; b -
\frac{1}{d-1})$

\hspace{1.4cm} {\tt else} {\bf Output} $(c, a, b)$

\medskip\noindent
As we can see, $c$ is a control line whose input value is returned unchanged.
The condition $c=1$ enables the movement of a quantity $\delta$ of energy from
the third to the second line.
Of course, this action is performed only if possible, that is, only if
$a \neq 1$ and $b \neq 0$ (equivalently, if the energy values associated to the
second and third line are not $\varepsilon_1$ and $\varepsilon_0$,
respectively).
If these conditions are not satisfied, or if $c \neq 1$, then the gate behaves
as the identity.
Starting from this description, for any integer $d \ge 2$ we can easily write
the truth table of the $d$--valued {\sc CUp} gate.

Analogously, the behavior of the complementary $(3,3,d)$--gate {\sc CDown} is:

\medskip
\hspace{1cm}{\bf Input}: $(c, a, b) \in L_d^3$

\hspace{1cm}{\tt if} $c = 1$

\hspace{1.4cm} {\tt then} {\bf Output} $(c, \; a - \frac{1}{d-1}, \; b +
\frac{1}{d-1})$

\hspace{1.4cm} {\tt else} {\bf Output} $(c, a, b)$

\medskip\noindent
Let us note that $\text{\sc CDown}(c,a,b)$ can be obtained from $\text{\sc CUp}
(c,a,b)$ (and vice versa) by exchanging the second and the third line before
and after the application of {\sc CUp}.

Figure \ref{fig:figures} (b) shows how, using the Boolean versions of {\sc CUp}
and {\sc CDown} gates, we can implement the Boolean Fredkin (controlled switch)
gate.
Since the Fredkin gate is functionally complete for Boolean logic, using only
two--valued {\sc CUp} and {\sc CDown} gates we can realize any Boolean circuit.
In principle these Boolean circuits, together with $d$--valued {\sc CUp}'s and
{\sc CDown}'s, can realize any conditional movement of energy, that is, any
conceivable computation that can be performed by $(n,m,d)$--gates.

It is clear that implementing a gate, be it conservative or not, using only
these primitives allows one to visualize the movement of energy between
different parts of the gate during a computation.
Such visualization may help us to optimize some aspects of the implementation
of the gate, namely, the amount of energy moved and the extension of energy
jumps.
As shown in \cite{leporati-phd}, such optimizations can be obtained by
splitting (if possible) a given $(N, M, d)$--gate $H$ into $k$ blocks, so that
its computation can be performed by an appropriate $(N/k, M/k, d)$--gate $G$
equipped with a storage unit of capacity $C$.
However, the minimization of the amount of energy moved between different
parts of $H$ during the computation is equivalent to the minimization of $C$,
and hence it constitutes an NP--hard problem, whose decision version is the
NP--complete problem {\sc ConsComp}.
This means that the reorganization of the internal machinery of $H$ to
optimize the movements of energy is considered a difficult problem.

Now let us turn to the quantum realization of {\sc CUp} and {\sc CDown}.
Generally, a quantum gate acts on memory cells that are $d$--level quantum
systems called \emph{qudits} (see \cite{cattaneo-leporati-leporini-qm} and
\cite{gottesman}).
A qudit is typically implemented using the energy levels of an atom or a
nuclear spin.
The mathematical description --- independent of the practical realization ---
of a single qudit is based on the $d$--dimensional complex Hilbert space
$\mathbb{C}^d$.
In particular, the truth values of $L_d$ are represented by the unit vectors
of the canonical orthonormal basis, called the \emph{computational basis} of
$\mathbb{C}^d$:
\begin{equation*}
   \ket{0} = \begin{bmatrix}
                1\\ 0 \\ \vdots \\ 0 \\ 0
             \end{bmatrix}, \hspace{0.5cm}
   \ket{\frac{1}{d-1}} = \begin{bmatrix}
                            0 \\ 1 \\ \vdots \\0\\ 0
                         \end{bmatrix}, \hspace{0.5cm}
   \cdots, \hspace{0.5cm}
   \ket{\frac{d-2}{d-1}} = \begin{bmatrix}
                              0 \\ 0 \\ \vdots \\ 1 \\ 0
                           \end{bmatrix}, \hspace{0.5cm}
   \ket{1} = \begin{bmatrix}
                0 \\ 0 \\ \vdots \\ 0 \\ 1
             \end{bmatrix}
\end{equation*}

A collection of $n$ qudits is called a \emph{quantum register} of size $n$.
It is mathematically described by the Hilbert space
$\otimes^n\C^d=\underbrace{\C^d\otimes\ldots\otimes\C^d}_{\mbox{\small $n$
times}}$.
An $n$--\emph{configuration} is a vector $\ket{x_1}\otimes\ldots\otimes
\ket{x_n} \in \otimes^n\C^d$, simply written as $\ket{x_1,\ldots,x_n}$, for
$x_i$ running on $L_d$.
An $n$--configuration can be viewed as the quantum realization of the
``classical'' pattern $(x_1,\ldots, x_n ) \in L_d^n$.
Let us recall that the dimension of $\otimes^n\C^d$ is $d^n$ and that the set
$\{\ket{x_1,\ldots,x_n}:x_i\in L_d\}$ of all $n$--configurations is an
orthonormal basis of this space, called the $n$--\emph{register computational
basis}.

Unlike the situation of the classical wired computer where voltages of a wire
go over voltages of another, in quantum realizations of classical gates
something different happens.
First of all, in this setting every gate must have the same number of input and
output lines (that is, they must be $(n, n, d)$--gates).
Each qudit of a given register configuration $\ket{x_1,\ldots,x_n}$ (quantum
realization of an input pattern) is in some particular quantum state
$\ket{x_i}$ and an operation $G:\otimes^n\C^d\mapsto\otimes^n\C^d$ is performed
which transforms this configuration into a new configuration
$G(\ket{x_1,\ldots,x_n}) = \ket{y_1,\ldots,y_n}$, which is the quantum
realization of an output pattern.
In other words, a quantum realization of an $(n, n, d)$--gate is a linear
operator $G$ that transforms vectors of the $n$--register computational basis
into vectors of the same basis.
The action of $G$ on a non--factorized vector, expressed as a linear
combination of the elements of the $n$--register basis, is obtained by
linearity.

The collection of all linear operators on $\C^d$ is a $d^2$--dimensional
linear space whose canonical basis is:
\[
  \left\{ E_{x,y}=\ket{y}\bra{x}\;:\; x,y\in L_d \right\}
\]
Since $E_{x,y}  \ket{x} = \ket{y}$ and $E_{x,y} \ket{z} = \mathbf{0}$
for every $z \in L_d$ such that $z \neq x$, this operator transforms the unit
vector $\ket{x}$ into the unit vector $\ket{y}$, collapsing all the other
vectors of the canonical orthonormal basis of $\C^d$ into the null vector.
For $i, j \in \{0, 1, \ldots, d-1\}$, the operator $E_{\frac{i}{d-1},
\frac{j}{d-1}}$ can be represented as an order $d$ square matrix having $1$ in
position $(j+1,i+1)$ and $0$ in every other position:
\[
   E_{\frac{i}{d-1},\frac{j}{d-1}} = \left(\delta_{r,j+1}\delta_{i+1,s}
      \right)_{r,s=1,2,\ldots,d}
\]

Each of the operators $E_{x,y}$ can be expressed, using the whole algebraic
structure of the associative algebra of operators, as a suitable composition
of creation and annihilation operators.
An alternative approach, that uses spin--creation and spin--annihilation
operators, is shown in \cite{cattaneo-leporati-leporini-qm}.
We recall that the actions of the \emph{creation} operator $a^\dag$ and of
the \emph{annihilation} operator $a$ on the vectors of the canonical
orthonormal basis of $\C^d$ are
\begin{align*}
   & a^\dagger \ket{\frac{k}{d-1}} = \sqrt{k+1} \ket{\frac{k+1}{d-1}}
        & \text{for $k \in \{0, 1, \ldots, d-2\}$} \\
   & a^\dagger \ket{1} = \mathbf{0}
\end{align*}
and
\begin{align*}
   & a \ket{\frac{k}{d-1}} = \sqrt{k} \ket{\frac{k-1}{d-1}}
        & \text{for $k \in \{1, 2, \ldots, d-1\}$} \\
   & a \ket{0} = \mathbf{0}
\end{align*}
respectively.
Hence, if denote by $A_{u,v}^{p,q,r}$ the expression
\[
   \underbrace{v \cdots v}_r \underbrace{v^\ast \cdots v^\ast}_q
      \underbrace{v \cdots v}_p u
\]
where $u,v \in \{a^\dag,a\}$, $v^\ast$ is the adjoint of $v$, and $p,q,r$ are
non negative integer values, then for any $i,j \in \{0,1,\ldots,d-1\}$ we can
express the operator $E_{\frac{i}{d-1},\frac{j}{d-1}}$ in terms of creation and
annihilation as follows:
\[
   E_{\frac{i}{d-1},\frac{j}{d-1}} =
   \begin{cases}
      \frac{ \sqrt{j!}}{(d-1)!}
      A_{a^\dag,a^\dag}^{d-2, d-1-j, 0} & \text{if $i = 0$} \\
      \frac{ \sqrt{j!}}{(d-1)!}
      A_{a,a^\dag}^{d-1, d-1-j, 0} & \text{if $i = 1$ and $j \ge 1$} \\
      \frac{ \sqrt{i!}}{(d-1)! \sqrt{j!}}
      A_{a^\dag,a^\dag}^{d-2-i, d-1, j} &
           \begin{minipage}[t]{5.9cm}
              if ($i = 1$, $j = 0$ and $d \ge 3$) or \\
              \indent\hspace{0.25cm} ($1 < i < d-2$ and $j \le i$)
           \end{minipage} \\
     \frac{ \sqrt{j!}}{(d-1)! \sqrt{i!}}
      A_{a,a}^{i-1, d-1, d-1-j} &
           \begin{minipage}[t]{5.9cm}
              if ($i = d-2$, $j = d-1$ and $d \ge 3$) \\
              \indent\hspace{0.25cm} or ($1 < i < d-2$ and $j > i$)
           \end{minipage} \\
     \frac{1}{ \sqrt{(d-1)! j! (d-1)}}
      A_{a^\dag,a}^{d-1, j, 0} & \text{if $i = d-2$ and $j \le d-2$} \\
      \frac{1}{ \sqrt{(d-1)! j!}}
      A_{a,a}^{d-2, j, 0} & \text{if $i = d-1$}
   \end{cases}
\]

Classical $(n,n,d)$--gates can be quantistically realized as sums of tensor
products of the operators $E_{x,y}$ as follows.
Let $x_1 x_2 \cdots x_n \mapsto y_1 y_2 \cdots y_n$ be a generic row of the
truth table of an $(n,n,d)$--gate.
For what we have said above, the operator $E_{x_1,y_1} \otimes E_{x_2,y_2}
\otimes \cdots \otimes E_{x_n,y_n}$ transforms the input configuration
$x_1 x_2 \cdots x_n$ into the output configuration $y_1 y_2 \cdots y_n$, and
collapses all the other input configurations of the $n$--register basis to the
null vector.
It is not difficult to see that if ${\cal O}_0, \ldots, {\cal O}_{d^n-1}$ are
the ``local'' operators associated to the $d^n$ rows of the truth table, then
the operator ${\cal O} = \sum_{i=0}^{d^n-1}{\cal O}_i$ is a quantum realization
of the $(n,n,d)$--gate.
Notice that the resulting operator ${\cal O}$ is not necessarily a unitary
operator.

Starting from the truth tables of the $d$--valued gates {\sc CUp} and
{\sc CDown} we can thus build the corresponding linear operators that realize
them.
For example, it is not difficult to see that the non--unitary linear operator
--- acting on the Hilbert space $\C^2 \otimes \C^2 \otimes \C^2$ --- which
realizes the Boolean {\sc CUp} gate is:
\begin{equation}
   {\rm Id} \otimes {\rm Id} \otimes {\rm Id} - c^\dagger c \otimes
       a a^\dagger \otimes b^\dagger b +
   ({\rm Id} \otimes a^\dagger \otimes b) (c^\dagger c \otimes a a^\dagger
   \otimes b^\dagger b)
\label{eq:CUp-formula}
\end{equation}
where ${\rm Id}$ is the identity operator of $\C^2$ and, for the sake of
clearness, we have written $c^\dagger$, $a^\dagger$, $b^\dagger$ (resp., $c$,
$a$, $b$) to denote the creation (resp., annihilation) operator of $\C^2$
applied onto the subspaces of $\C^2 \otimes \C^2 \otimes \C^2$ corresponding
to the first, second and third input, respectively.
In fact, the gate behaves as the identity if the input pattern
$\ket{x_c,x_a,x_b}$ is different from $(1,0,1)$, since in these cases
$(c^\dagger c \otimes a a^\dagger \otimes b^\dagger b)\ket{x_c,x_a,x_b} =
\mathbf{0}$, the null vector of $\C^2 \otimes \C^2 \otimes \C^2$.
On the other hand $(c^\dagger c \otimes a a^\dagger \otimes b^\dagger b)
\ket{1,0,1} = \ket{1,0,1}$, hence the first two terms of \eqref{eq:CUp-formula}
dissapear and the operator $({\rm Id} \otimes a^\dagger \otimes b)$ is applied
on $\ket{1,0,1}$, giving $\ket{1,1,0}$ as required.

In a completely analogous way we can see that the non--unitary linear operator
which realizes the Boolean {\sc CDown} gate is:
\[
   {\rm Id} \otimes {\rm Id} \otimes {\rm Id} - c^\dagger c \otimes
       a^\dagger a \otimes b b^\dagger +
   ({\rm Id} \otimes a \otimes b^\dagger) (c^\dagger c \otimes a^\dagger a
   \otimes b b^\dagger)
\]

Let us note that the use of creation and annihilation operators allows for
different physical implementations.
For example, we can view computation not only as a conditional movement of
energy but also as a conditional movement of particles between systems that may
contain at most $d-1$ of particles.
Alternatively, we can view computation as a sequence of conditional switches of
the value of the $z$ component of the angular momentum of microscopical
physical systems, using spin--creation and spin--annihilation instead of
creation and annihilation operators \cite{cattaneo-leporati-leporini-qm}.

\section{Conclusions and directions for future work}

In this paper we have proposed the first steps towards a theory of conservative
computing, where the amount of energy associated to the data which are
manipulated during the computations is preserved.

The first obvious extension of our model is to take into account the energy
used to perform computations, that is, to transform input values into output
values.
A first idea is to consider some additional \emph{power source} input lines and
\emph{dissipation} output lines.
Power source lines are fixed to a constant value from $L_d$ (usually $1$), and
absorb energy from the environment.
This energy is entirely consumed during the computation, whereas all the energy
associated to the input pattern is returned by the output pattern.
On the other hand, dissipation lines are used to model the release of energy
into the environment; hence, their value is simply discarded.
Conservative gates constitute a special case in our framework, where there are
neither power source nor dissipation lines (under the hypothesis that we do not
take into account the energy needed to perform the computation).

Since perfect conservation of energy can be obtained only in theory, a second
possibility for future work could be to relax the conservativeness constraint
\eqref{eq:conservativegate}, by assuming that the amount of energy dissipated
during a computation step is not greater than a fixed value.
Analogously, we can suppose that if we try to store an amount of energy that
exceeds the capacity of the gate then the energy which cannot be stored is
dissipated.
In such a case it should be interesting to study trade-offs between the amount
of energy dissipated and the hardness of the corresponding modified
{\sc ConsComp} problem.

Finally, it remains to study how to theoretically model and physically realize
gates equipped with an internal storage unit.
Here we just observe that, from a theoretical point of view, it seems
appropriate to consider this kind of gates as finite state automata, by viewing
the energy levels of the storage unit as their states.

\end{document}